\documentclass[fleqn,usenatbib]{mnras}

\usepackage{newtxtext,newtxmath}

\usepackage[T1]{fontenc}

\DeclareRobustCommand{\VAN}[3]{#2}
\let\VANthebibliography\thebibliography
\def\thebibliography{\DeclareRobustCommand{\VAN}[3]{##3}\VANthebibliography}

\usepackage{graphicx}	
\usepackage{amsmath}	
\usepackage{derivative}
\usepackage[dvipsnames]{xcolor}
\usepackage{ulem}

\newcommand{\aimf}{\alpha_{\text{IMF}}}

\newcommand{\inprepp}[1]{(#1 \textit{in prep.})}

\title[Stellar mass from microlensing]{Stars as cosmic scales: measuring stellar mass with microlensed supernovae}

\author[Weisenbach et al.]{Luke Weisenbach,$^{1,}$\thanks{E-mail: weisluke@alum.mit.edu}
Thomas Collett,$^{1}$
Wolfgang Enzi,$^{1}$
Lindsay Oldham,$^{1}$
Ana Sainz de Murieta$^{1}$\\
{$^{1}$Institute of Cosmology and Gravitation, University of Portsmouth, Burnaby Road, Portsmouth, PO1 3FX, UK}
} 

\date{Accepted XXX. Received YYY; in original form ZZZ}

\pubyear{2024}

\begin{document}
\label{firstpage}
\pagerange{\pageref{firstpage}--\pageref{lastpage}}
\maketitle

\begin{abstract}

Gravitational microlensing is a unique probe of the stellar content in strong lens galaxies. Flux ratio anomalies from gravitationally lensed supernovae (glSNe), just like lensed quasars, can be used to constrain the stellar mass fractions at the image positions. Type Ia supernovae are of particular interest as knowledge of the intrinsic source brightness helps constrain the amount of (de)magnification from the macromodel predictions that might be due to microlensing. In addition, the presence or absence of caustic crossings in the light curves of glSNe can be used to constrain the mass of the microlenses. We find that a sample of 50 well-modeled glSNe Ia systems with single epoch observations at peak intrinsic supernova luminosity should be able to constrain an average stellar mass-to-light ratio to within $\sim 15\%$. A set of systems with light curve level information providing the location (or absence) of caustic crossing events can also constrain the mass of the microlenses to within $\sim 50\%$. Much work is needed to make such a measurement in practice, but our results demonstrate the feasibility of microlensing to place constraints on astrophysical parameters related to the initial mass function of lensing galaxies without any prior assumptions on the stellar mass.

\end{abstract}




\section{Introduction}
\label{sec:intro}

Precisely how the matter content of galaxies is split between baryons and dark matter is a complex problem. Various methods for measuring the relative contributions of each have been explored, ranging from stellar population synthesis (SPS) \citep{2013ARA&A..51..393C}, kinematics \citep{2016ARA&A..54..597C}, strong gravitational lensing \citep{2024SSRv..220...87S}, and combinations thereof. In the case of the former, a fundamental input required for fitting the spectral energy distribution of a galaxy is the Initial Mass Function (IMF) of the stellar population. In tandem with the star formation rate, the IMF controls the Present Day Mass Function (PDMF) and therefore the present day total stellar mass. While there is increasing evidence for variations in the IMF among galaxies \citep{2020ARA&A..58..577S}, measurements of stellar mass across a variety of galactic parameters such as age, metallicity, and velocity dispersion may lend insight into the mechanisms by which the stellar populations of galaxies evolve. 

In the case of gravitational lensing, the quantity that is best constrained by data is the total mass within the Einstein radius. Assuming functional forms for the dark and/or baryonic components and combining lensing with stellar kinematics can provide a measurement of the stellar mass \citep{2004ApJ...611..739T, 2012ApJ...752..163S, 2015ApJ...800...94S} while breaking some degeneracies. Results for different samples of lenses favor either Chabrier \citep{2019A&A...630A..71S} or Salpeter IMFs \citep{2011MNRAS.417.3000S, 2012ApJ...753L..32S, 2021MNRAS.503.2380S}, though constraints depend on the functional forms used as well as how the stellar mass is parametrized, with some tentative hints of trends with velocity dispersion \citep{2010ApJ...709.1195T}.

In all cases, the (dis)agreement with specific IMFs relies on comparing the inferred stellar mass from lensing to that from SPS. Along with the shape and slope of the IMF, such results are sensitive to the low mass cutoff used in the IMF -- which can significantly change the stellar mass-to-light ratio $\Upsilon_\star$ \citep{2013MNRAS.436..253B, 2017ApJ...837..166C, 2017ApJ...845..157N}. While lensing is sensitive to the total two dimensional projected mass, a portion of that mass is due to stellar remnants, planets, and stars near the hydrogen burning limit which do not contribute significantly (if at all) to the measured light \citep{2018ASPC..514...79S}. They do, however, source small scale gravitational lensing deflections.

When a background source is appropriately aligned behind a foreground galaxy, the gravitational potential of the galaxy distorts and (de)magnifies the image of the source, at times creating multiple images if the alignment is fortuitous. The stars within the lensing galaxy also act as lenses in their own right \citep{1979Natur.282..561C, 1981ApJ...244..756Y, 1986ApJ...301..503P}, further perturbing the multiple images when the source is of a sufficiently small size. Known as microlensing, this effect is sensitive to all compact objects in the lens galaxy along the line of sight. This makes it a unique probe of the stellar mass fraction, and therefore the PDMF, that does not depend on assumptions about the stellar mass-to-light ratio or the stellar mass profile -- just the total mass profile, which is typically well constrained by the separations between macroimages. 

For the majority of strong lensing history, the only sources of sufficient compactness to allow for such microlensing to occur have been quasars \citep{1979Natur.279..381W, 1989AJ.....98.1989I, 2010GReGr..42.2127S}. However, recent discoveries of gravitationally lensed supernovae (glSNe) \citep{2017Sci...356..291G, 2023NatAs...7.1098G} have opened a new door for microlensing studies. Microlensing is as ubiquitous for lensed supernovae \citep{2006ApJ...653.1391D, 2024SSRv..220...13S} as it is for lensed quasars \citep{1995ApJ...443...18W, 2024SSRv..220...14V}, and given the wealth of glSNe expected to be discovered by the Vera Rubin Observatory (LSST) in the next decade \citep{2019ApJS..243....6G, 2019MNRAS.487.3342W, 2023MNRAS.526.4296S, 2024MNRAS.531.3509A} it is worth understanding precisely what information can be gleaned from them. 

Much work has already been done on the utility of glSNe, specifically Type Ia supernovae (glSNe Ia), as probes of cosmology through time delay cosmography \citep{2018ApJ...855...22G, 2018MNRAS.478.5081F, 2019A&A...631A.161H, 2019ApJ...876..107P, 2021A&A...646A.110H}. Other works have examined how glSNe can be used to constrain the total mass profile of the lens galaxy \citep{2020MNRAS.496.3270M, 2022A&A...662A..34D}; glSNe Ia shine here as their `standardizable candle' nature might permit breaking the mass-sheet degeneracy \citep{1985ApJ...289L...1F, 2014A&A...564A.103S} and testing lens modeling systematics -- though only if microlensing is negligible compared to the intrinsic scatter in the luminosity of the Ia population \citep{2018MNRAS.478.5081F, 2024MNRAS.531.4349W}. In the cases where microlensing is not negligible however, the problem can be flipped around: rather than using glSNe Ia to probe cosmology, reasonable assumptions on cosmological parameters instead turn them into probes of the stellar content of the lensing galaxy. 

Various microlensing constraints on the IMF have been made using lensed quasars in both the X-ray and optical (see \citealt{2024SSRv..220...14V} for a review). Constraints are typically on either some constant stellar mass fraction \citep{2009ApJ...697.1892P, 2009ApJ...706.1451M} or a constant stellar mass-to-light ratio \citep{2014ApJ...793...96S}. In the case of the former, the assumption of constancy seems roughly appropriate if the images are all at similar distances from the center of the lens. The latter approach may be more robust, though mass-to-light gradients in the lens galaxy could alter results. More generally however, what microlensing allows us to \textit{actually} measure is the stellar mass-to-light ratio of the PDMF at the image positions. Precisely how such measurements then relate to constraints on the IMF throughout the lens galaxy depends on assumptions one may or may not wish to make or test.

In addition to constraining the stellar mass fraction, microlensing also probes the typical mass of a compact object at the image locations. Sensitivity to mass at the lower end of the mass spectrum is important for constraining the low mass cutoff of the IMF, which along with the shape of the mass spectrum determines the mean mass \citep[see the range in, e.g., Table 2 of][for various IMFs]{2011ApJ...726...27P}. The finite size of the source is an extremely important nuisance factor in such measurements \citep{2006ApJ...645..835L}, but joint constraints of quasar source size and typical stellar mass have been made \citep{2017ApJ...836L..18M, 2019ApJ...885...75J}. 

Supernovae offer a host of improvements over quasars that are beneficial. They 1) evolve over shorter timescales, requiring less observational investment in light curve monitoring; 2) fade away, permitting better followup of the lens and host galaxy light for modeling; and 3) have better understood sizes, luminosities, and light curve evolution \citep{1993ApJ...413L.105P, 2021ApJ...923..265K}, which allows for easier microlensing analyses. \citet{2006ApJ...653.1391D} noted that ``...the amount of structure in the [microlensed] light curves is a sign that more effort should be exerted to figure out how to analyze lensed SN light curves to determine properties of the stellar population.'' The purpose of this work is to exert just such an effort on constraining 1) the fraction of matter in the form of compact objects and 2) the typical mass of those compact objects, at the positions of lensed supernovae images. 

In Section~\ref{sec:theory}, we review how microlensing can be used to constrain the stellar mass fraction and the typical mass of compact objects. In Section~\ref{sec:sims}, we discuss the mock data used in this work. In Section~\ref{sec:constraining_aimf}, we explore how well we can recover the stellar mass fractions from a population of glSNe Ia. In Section~\ref{sec:constraining_aimf_m}, we further examine how well we can jointly constrain the stellar mass fraction and the microlens mass. We discuss some limitations of our work in Section~\ref{sec:caveats}, and present conclusions and discussions in Section~\ref{sec:conclusions}.

\section{Microlensing constraints on stellar mass parameters}
\label{sec:theory}

In this section, we provide an overview of why microlensing can be used to constrain the stellar mass fraction and the typical mass of compact objects in the lens galaxy.

\subsection{Constraining the stellar mass fraction}
\label{subsec:constraining_aimf}

\begin{figure*}
    \centering
    \includegraphics[width=\columnwidth]{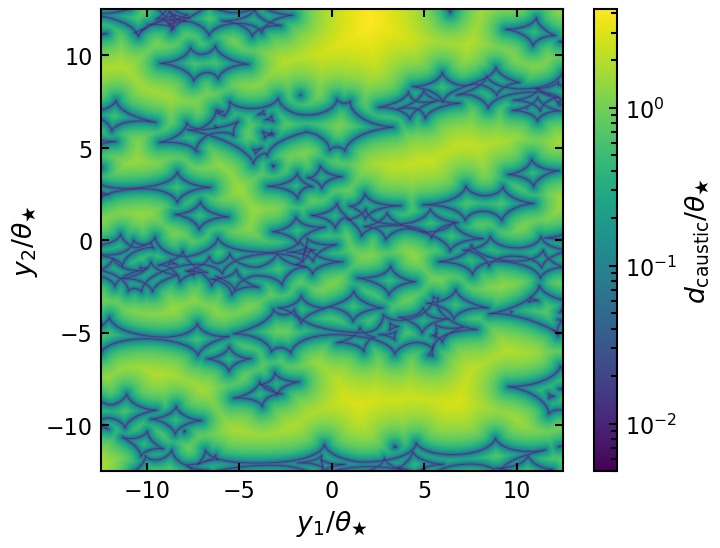}
    \hspace{15pt}
    \includegraphics[width=\columnwidth]{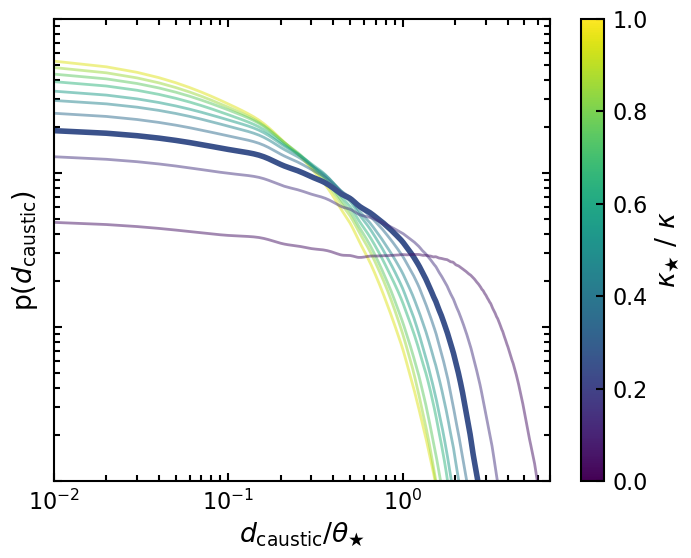}\\
    \includegraphics[width=\columnwidth]{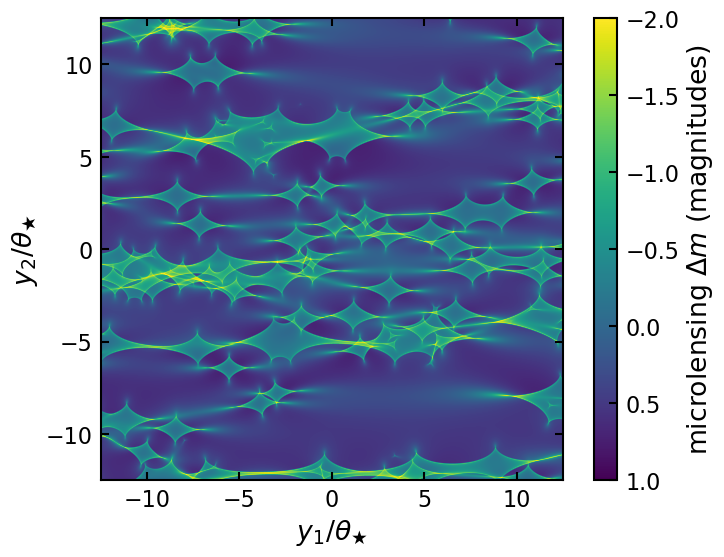}
    \hspace{15pt}
    \includegraphics[width=\columnwidth]{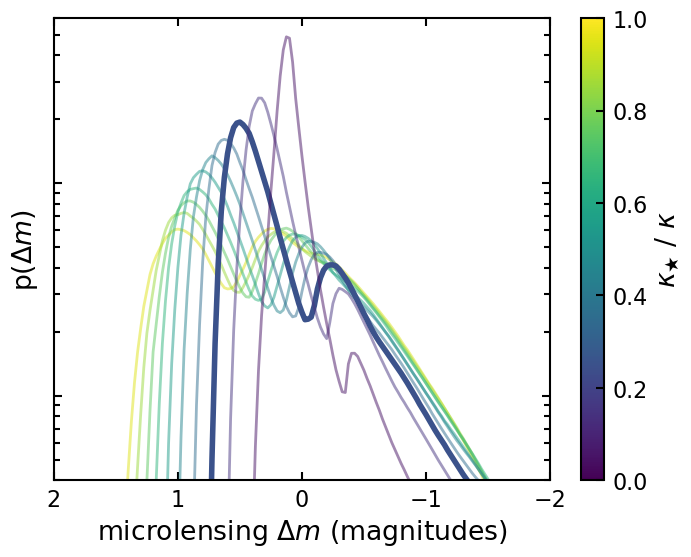}
    \caption{Visualization of the distance to the nearest caustic $d_{\text{caustic}}$ (top left) and the microlensing (de)magnification $\Delta m$ relative to the macromodel (bottom left) as a function of position in the source plane for $\kappa=\gamma=0.4$. The probability distributions of $d_{\text{caustic}}$ and $\Delta m$ are shown in the right column. The thick lines show the specific distributions for the value of $\kappa_\star/\kappa$ of the left column, while the thinner opaque lines show how the distributions change with and are highly sensitive to the ratio $\kappa_\star / \kappa$. We show examples of the joint distributions of $d_{\text{caustic}}$ and $\Delta m$ in Figure~\ref{fig:d_caustic_mags} which highlight the strong correlations between the variables.}
    \label{fig:d_caustic_delta_m}
\end{figure*}

The idea behind measuring the stellar mass fraction from microlensing uses the fact that the distribution of microlensing magnification is highly sensitive to the value of the stellar mass fraction at the image locations, as shown in the bottom right of Figure~\ref{fig:d_caustic_delta_m} for a point source. Some values of (de)magnification strongly favor particular stellar mass fractions, and can even rule out ranges of stellar mass fractions as impossible. In general, one can use the flux ratios between pairs of images to calculate the likelihood of the stellar convergence $\kappa_\star$ for each image based on the observed magnification ratios and the microlensing magnification distributions \citep{2002ApJ...580..685S}. 

A necessary ingredient to make such measurements however is the value of the total convergence $\kappa$ at the image locations. In general $\kappa$ is potentially difficult to constrain due to the mass-sheet degeneracy \citep{1985ApJ...289L...1F}, the choice of how to parametrize the gravitational potential of the lens, and uncertainties on cosmological parameters of the universe such as the Hubble constant $H_0$. There are a few ways such concerns can be mitigated. Spatially resolved kinematics measurements of the lens galaxy can be used to map out the gravitational potential and break the mass-sheet degeneracy \citep{2018MNRAS.473..210S, 2023A&A...675A..21Y}. Alternatively, if one assumes a cosmology, the time delay(s) between images break the mass-sheet degeneracy as well. Einstein radii can be measured to a few percent, while time delays for lensed supernovae can be measured to within a few days \citep{2018ApJ...855...22G, 2021A&A...646A.110H, 2021A&A...653A..29B, 2022A&A...658A.157H}; picking one's favored value of $H_0$ from current early or late time measurements \citep{2020A&A...641A...6P, 2022ApJ...934L...7R} or some less constrained combination of the two then permits $\kappa$ to be known to within a few percent -- sufficient for our purposes.

An important nuisance factor in microlensing measurements of stellar mass fractions is the size of the source relative to the characteristic length scale (Einstein radius) of the microlenses $\theta_\star$. Supernovae provide a boon over lensed quasars in that the size of the source is better constrained. The X-ray emitting regions of quasars are the most ideal for microlensing measurements as they come from a compact region which can generally be considered as point-like, or at least much smaller than the region which emits in the visible or larger wavelengths \citep{2006ApJ...648...67P, 2007ApJ...661...19P}. Supernovae also start off orders of magnitude smaller than the Einstein radii of the microlenses; for representative source and lens redshifts of 0.4 and 0.8 (respectively) with microlenses of mass $0.3M_\odot$, a supernova expanding at $10^4$ km/s reaches a size of $\sim0.2\theta_\star$ after 50 rest-frame days. Treating supernovae as point-like is an acceptable approximation for some microlensing purposes (compare \citealt{2018MNRAS.478.5081F} and \citealt{2021ApJ...922...70W}), easing the need to jointly model the size and profile of the source. Should one wish to be more precise, the expansion speed of supernovae can be determined from spectral lines \citep{1974ApJ...193...27K, 1996ApJ...466..911E, 2005A&A...439..671D, 2012MNRAS.419.2783T}, providing a prior on the source size that can be used when modeling. Theoretical supernovae explosion models can also provide half-light radii as a function of wavelength and time \citep{2018ApJ...855...22G, 2019A&A...631A.161H}, which will likely permit multispectral observations and modeling to provide more stringent constraints than this work in the future. 

Furthermore, Type Ia supernovae constrain an additional unknown -- the intrinsic brightness of the source. Having an informed prior on the source brightness constrains the microlensing (de)magnification\footnote{if we ignore millilensing} and is important for reducing the number of gravitationally microlensed systems required to make interesting conclusions. 

\subsection{Constraining the microlens mass}
\label{subsec:constraining_m}

While single epoch flux ratio anomaly measurements can be used to constrain stellar mass fractions, light curves provide more than ample information to make such measurements as well \citep{2004ApJ...605...58K}; each of the methods have their merits and drawbacks (see \citealt{2024SSRv..220...14V} for a review). The time evolution of light curves additionally make it easier to constrain the source size and the microlens mass from temporal correlations. We do not perform full light curve modeling in this work, but we will instead investigate just how well microlens mass can be constrained with simple information taken from light curves.

It is well known that the physical size of the source relative to $\theta_\star$ is one of the main factors that drives microlensing variability -- moreso than its projected two dimensional luminosity profile \citep{2005ApJ...628..594M, 2019MNRAS.486.1944V}. The degeneracy between source size and microlens mass (as $\theta_\star\propto\sqrt{m_\star}$) is typically broken through different priors on each. As supernovae expand at rates of tens to a few tens of thousands of km/s \citep{2012MNRAS.419.2783T, 2018ApJ...858..104Z}, providing a known physical scale that is better constrained than the varying sizes of quasar emission regions, we might therefore expect to learn most easily about the average mass of the microlenses. The situation is somewhat complicated by the fact that a spectrum of masses from a PDMF introduces a range of length scales which may be relevant for microlensing at different points in time. We limit ourselves in this work to the simplest scenario where all the microlenses have the same mass, though we discuss the implications of a mass spectrum in Section~\ref{sec:caveats}. We fully explore how well this single mass can be constrained with fairly simple considerations in Section~\ref{sec:constraining_aimf_m}, but we provide some motivation for such constraints below.

\begin{figure}
    \centering
    \includegraphics[width=\columnwidth]{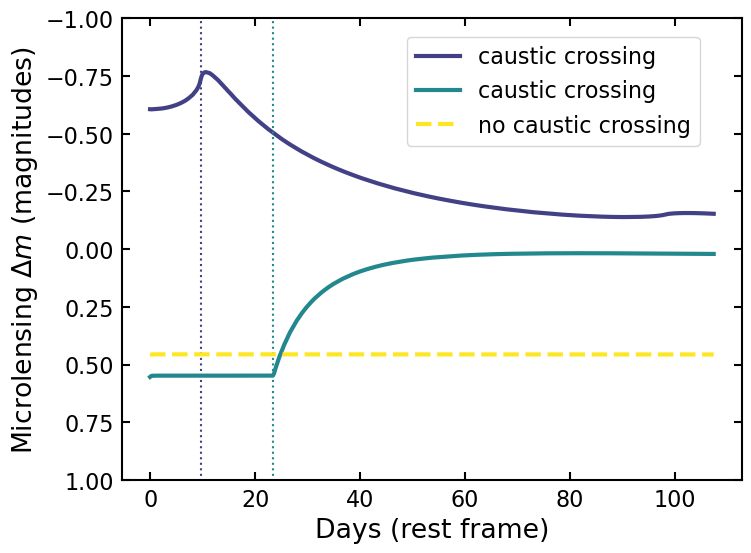}\\
    \vspace{15pt}
    \includegraphics[width=\columnwidth]{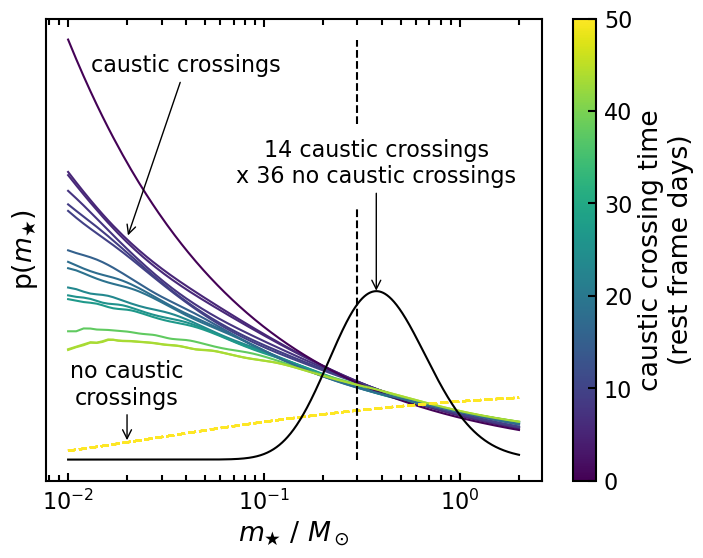}
    \caption{Top: Sample light curves showing the temporal evolution of microlensing magnification as a supernova photosphere expands. Caustic crossings can induce substantial magnification changes that should be identifiable upon removing the intrinsic Ia signal. Vertical dotted lines mark the time of caustic crossing. Bottom: Posteriors (colored curves) of $m_\star$ calculated from the caustic crossing times for 50 different microlensing light curves from the same macroimage, and their product (black curve). The vertical dashed black line denotes the true value ($m_\star=0.3M_\odot$) used when creating the mock data. Light curves which do not cross caustics all have the same dashed yellow posterior. Posteriors from individual light curves do not well constrain the mass, but highlight the dependence on the caustic crossing time.}
    \label{fig:mass_posteriors}
\end{figure}

Unlike quasars which have a fixed size and move through the microcaustic network, supernovae have a fixed position and expand in size through the microcaustic network. Occasionally as the supernova expands, the photosphere will cross a microcaustic, inducing either a sharp increase or a peak followed by a gradual decrease in the light curve which should be identifiable relative to the known light curve evolution of glSNe Ia (see Figure~\ref{fig:mass_posteriors}). This allows us to measure the microlens mass from the time (or lack of time) at which the first caustic crossing event occurs in the light curve, so long as we know the expansion velocity of the supernova. 

Using simulations that provide the locations of the microcaustics, we can determine the distance to the nearest caustic $d_\text{caustic}$ as a function of source position (shown in Figure~\ref{fig:d_caustic_delta_m}) in units of $\theta_\star$. A large microlens mass means an increased separation (in physical units) between caustics, decreasing the likelihood a caustic crossing should occur as the supernova expands. Conversely, a small microlens mass means a decreased separation between caustics and an increased chance of a caustic crossing. This can be visualized by considering the map of $d_\text{caustic}$ in the top left of Figure~\ref{fig:d_caustic_delta_m}; a specific value of $m_\star$ merely sets the length scale for the axes in some physically meaningful units. The probability distribution $p(d_\text{caustic}|\theta_\star)$ can therefore be used to constrain $m_\star$ from caustic crossing events.

To highlight this, we create 50 realizations of microlensing light curves for a single macroimage with a fixed stellar surface mass density using the procedures outlined later in Section~\ref{sec:sims} and assume that we can precisely measure the time (or absence) of caustic crossings in the light curves. For an individual light curve, the probability of mass $m_\star$ given the caustic crossing time $t_{\text{caustic}}$ is \begin{equation}
    p(m_\star|t_{\text{caustic}}) \propto p(t_{\text{caustic}}|m_\star)p(m_\star)
\end{equation} We take $p(m_\star)$ to be a uniform prior between $0.01$ and $2$ $M_\odot$. The value of $m_\star$ directly depends on $\theta_\star$, while the value of $d_\text{caustic}$ directly depends on $t_\text{caustic}$ assuming a known expansion velocity for the supernova. The likelihood $p(t_{\text{caustic}}|m_\star)$ is then related to $p(d_{\text{caustic}}|\theta_\star)$, and we can therefore directly translate the distribution of distance-to-caustics into a posterior on the mass given some observed time (or absence) of caustic crossing.

Posteriors for the 50 light curves are shown in Figure~\ref{fig:mass_posteriors}. As expected, light curves which have caustic crossings at early times heavily favor low mass microlenses. As the time of caustic crossing increases, the probability of low mass decreases, while the probability of high mass increases. If no caustic crossing is seen within the time of observations, the posterior for the mass uses the integrated probability from the maximum observation time to the maximum distance to a caustic. While an individual light curve does not have much useful constraining power, the final combination of 50 light curves does -- with a peak at $m_\star\approx 0.38M_\odot$, though the distribution is skewed with $m_\star=0.54^{+0.44}_{-0.23} M_\odot$ (68\% confidence interval). The posterior has scattered high from truth ($m_\star=0.3M_\odot$) in this example due to Poisson noise in the number of the 50 curves that crossed a caustic.

\section{Mock data}
\label{sec:sims}

We use a set of simulated glSNe Ia from \citet{2023MNRAS.526.4296S} as our mock data set, with a cut on image separation $>0.8''$ (i.e. resolvable). We also restrict ourselves to source redshifts for which the supernovae would be observable at peak in the absence of lensing, to guarantee that at least one image is observed\footnote{This requires a peak brightness $\gtrsim 24$ mag for LSST.}. The catalog contains the macromodel parameters for simulated lenses and the positions and convergence values for the images of the sources. We model the stellar mass at the image locations with an elliptical de Vaucouleurs profile for the light aligned with the SIE potential of the lens \citep{2019MNRAS.483.5583V}. Effective radii for the lenses are calculated from their velocity dispersions using the scaling relations in \citet{2009MNRAS.394.1978H}. We then determine the mock stellar mass fractions assuming a Chabrier IMF, using fits for the dark matter fraction within half the effective radius from \citet{2010ApJ...724..511A}. 

We simulate microlensing magnification maps for the image parameters, using a mass of $m_\star=0.3M_\odot$ for all the microlenses. A true microlensed magnification light curve is generated for each image assuming a uniform disk for the supernova profile that is expanding at $10^4$ km/s in the network of microcaustics. We take the light curves to extend up to 50 days in the rest frame. We simulate the lensed magnitude in the $i$ band using the the microlensed magnification and the spectroscopic templates of \citet{2007ApJ...663.1187H} for the unlensed brightness, adding a 0.1 mag Gaussian scatter to mimic the intrinsic scatter present in the brightness of the Ia population after standardization. We add a further 0.01 mag uncertainty on the magnification to simulate observational photometric uncertainties. 

In Section~\ref{sec:constraining_aimf}, we use the brightness of the lensed images at peak intrinsic luminosity. Essentially, this assumes that we discover the glSNe before the first image reaches peak, and have observations at peak for both images. In Section~\ref{sec:constraining_aimf_m}, we will use some information from the light curves, which again assumes discovery of the first image before it reaches peak intrinsic luminosity, along with adequate monitoring of it and subsequent images as the light curve evolves.

We note that while the catalog contains all images, the number of observed images for a double can be 1 if the second (saddlepoint) image is too demagnified to be observed. Similarly, a quad may have less than 4 detected images. We use sample sizes of 50 or 200 systems in the following sections; there was one quad in the sample of 50, and 3 in the sample of 200. While doubles in general do not outnumber quads by the ratio suggested here \citep{2023MNRAS.526.4296S}, our requirement that the images be well separated reduces the number of quads in the catalog that are usable for our purposes. Quadruply imaged glSNe Ia from small Einstein radii systems such as the two currently known galaxy-scale cases \citep{2017Sci...356..291G, 2023NatAs...7.1098G} have more stringent observation requirements such as adaptive optics or space based followup, but pose an excellent way to probe the stellar content in small mass galaxies \citep{2020MNRAS.496.3270M, 2025arXiv250101578A}.

\section{Constraining the stellar mass fraction}
\label{sec:constraining_aimf}

In this section, we discuss the methodology for constraining the stellar mass fraction at the image locations from single epoch observations, with results for how well we can recover the input values from our simulations.

\subsection{Methodology}

We assume that, for each system, we can perfectly measure the convergence $\kappa$ at the image locations as well as the effective radius and ellipticity of the lens. Our goal then is to determine how well we can recover the stellar mass fractions based on the microlensing (de)magnifications. While microlensing measures $\kappa_\star$, the underlying physical phenomenon which drives the value of $\kappa_\star$ is the stellar mass-to-light ratio $\Upsilon_\star$. Following \citet{2014ApJ...793...96S}, we take the free parameter of the problem to be the ratio of the value of $\kappa_\star$ measured from microlensing to a fiducial $\kappa_{\star, \text{IMF}}$ calculated from a specific IMF, i.e. an IMF mismatch $\alpha_{\text{IMF}}$ \citep{2010ApJ...709.1195T} (which is equivalent to letting the free parameter be $\Upsilon_\star$). We use Chabrier as our fiducial IMF as well, and denote the free parameter as \begin{equation}
    \aimf = \frac{\kappa_\star}{\kappa_{\star, \text{Chabrier}}}
\end{equation} This means that we expect to recover a value of $\aimf=1$, and the goal is to see just how well we are able to do so with varying sample sizes and uncertainties.

For the macromodel convergence $\kappa$ and shear $\gamma$ of each image, we create microlensing magnification maps using inverse polygon mapping \citep{2006ApJ...653..942M, 2011ApJ...741...42M} combined with the fast multipole method \citep{1987JCoPh..73..325G} for approximating the deflection angle of distant microlenses \citep{2022ApJ...941...80J}, implemented on a GPU to speed up calculations by orders of magnitude \inprepp{Weisenbach} and make the analyses tractable. We sample the smooth matter fraction $s = 1 - \frac{\kappa_\star}{\kappa}$ in discrete steps of $\Delta s = 0.02$ and average multiple maps for each $(\kappa, \kappa_\star)$ pair to reduce noise in the final microlensing magnification distributions. 

For each system, we are interested in the posterior \begin{equation}
    p(\aimf | D)\propto p(D|\aimf)p(\aimf)
\end{equation} of $\aimf$ given some data $D$ which consists of the observed brightnesses (in magnitudes) of the images in the $i$ band. There are a variety of other parameters which must be considered and marginalized over however. Three are common to each image: the microlens mass $m_\star$, the unlensed source brightness (in magnitudes) at peak intrinsic luminosity $m_\text{s}$, and the source half-light radius $R_{1/2}$. We are therefore more generally interested in the posterior \begin{equation}
    \begin{aligned}
        &p(\aimf, m_\star, m_\text{s}, R_{1/2}|D)\\
        &=p(D|\aimf,m_\star,m_\text{s},R_{1/2})\\
        &\times p(\aimf,m_\star)p(m_\text{s},R_{1/2})
    \end{aligned}
\end{equation} where we discuss the likelihood and the priors individually below.

The likelihood $p(D|\aimf,m_\star,m_\text{s},R_{1/2})$ is equal to the product of the likelihoods of the individual images, as microlensing introduces independent (de)magnifications to each of them, \begin{equation}
    p(D|\aimf,m_\star,m_\text{s},R_{1/2}) = \prod_j p(D_j|\aimf,m_\star,m_\text{s},R_{1/2})
\end{equation} where the index $j$ runs over the number of images. The likelihood of an individual image must further account for an additional nuisance factor: the (unknown) true microlensed image brightness (in magnitudes) $m_j$, \begin{equation}
    \begin{aligned}
        &p(D_j, |\aimf,m_\star,m_\text{s},R_{1/2})\\
        &= \int p(D_j, m_j|\aimf,m_\star,m_\text{s},R_{1/2})\odif{m_j}\\
        &= \int p(D_j|m_j)p(m_j|\aimf,m_\star,m_\text{s},R_{1/2})\odif{m_j}        
    \end{aligned}
\end{equation} as the observed image brightness $D_j$ only depends on the parameters of interest hierarchically through $m_j$. We take the likelihood for each image $p(D_j|m_j)$ to be a Gaussian centered at $m_j$ with width equal to the observational uncertainty $\sigma_j$ on the observed brightnesses. The true image brightnesses $m_j$ are deterministic from $m_\text{s}$ and the microlensing magnifications $\mu_j$, but values of $\mu_j$ depend on the stellar convergences $\kappa_{\star,j}$, the microlens mass $m_\star$, and the source half-light radius $R_{1/2}$. The values of $\kappa_{\star,j}$ are deterministic from $\aimf$ however. We therefore use the microlensing magnification probability distributions (expressed in magnitudes) as $p(m_j|\aimf,m_\star,m_\text{s},R_{1/2})$. For each system then, we have the posterior\begin{equation}
\begin{aligned}
    &p(\aimf, m_\star, m_\text{s}, R_{1/2} | D)\\
    &= \prod_j\int p(D | m_j, \sigma_j) p(m_j|\aimf, m_\star, m_\text{s}, R_{1/2})\odif{m_j}\\
    &\times p(\aimf, m_\star)p(m_\text{s},R_{1/2})
\end{aligned}
\end{equation}

A joint prior for the source $p(m_\text{s},R_{1/2})$ may in general come from Type Ia supernova explosion models. For our purposes, we assume $m_\text{s}$ and $R_{1/2}$ are independent. In that case, we take the prior for $m_\text{s}$ to be a Gaussian centered on the \citet{2007ApJ...663.1187H} template magnitude for the source redshift; we will explore the effects of various levels of uncertainties in the source brightness. The prior on $R_{1/2}$ should in general come from measurements of the expansion velocity using spectra; in this work we will further assume the size is known exactly.

The prior $p(\aimf, m_\star)$ might come from SPS models; as we are interested in how well microlensing can constrain the parameters with no other information, we take a uniform prior subject only to the constraint that the $\kappa_{\star,j}$ are positive and can never be greater than the total convergence $\kappa_j$\footnote{The tophat priors that ensure $\kappa_{\star,j} / \kappa_j$ lie between 0 and 1 correspond to tophat priors between 0 and a different maximum $\aimf$ for each image.}.

We make two simplifying assumptions to reduce the dimensionality of the parameter space. First, while our mock data were created using expanding supernova profiles, we model them here as point sources to avoid the computational burden of convolving magnification maps. This removes $R_{1/2}$ and any sensitivity to $m_\star$; we will return to this approximation in Section~\ref{sec:constraining_aimf_m}. Second, we choose to ignore the small observational errors in the image magnitudes, as they are expected to be of order $0.01$ mag for LSST -- an order of magnitude smaller than the scatter in the intrinsic brightness of the Ia population. This essentially assumes that the observed magnitudes $D_j$ are equal to the true microlensed magnitudes $m_j$, removing the integrals over $m_j$ within the likelihood and absorbing the small observational uncertainties into the intrinsic luminosity prior.

\begin{figure}
    \centering
    \includegraphics[width=\columnwidth]{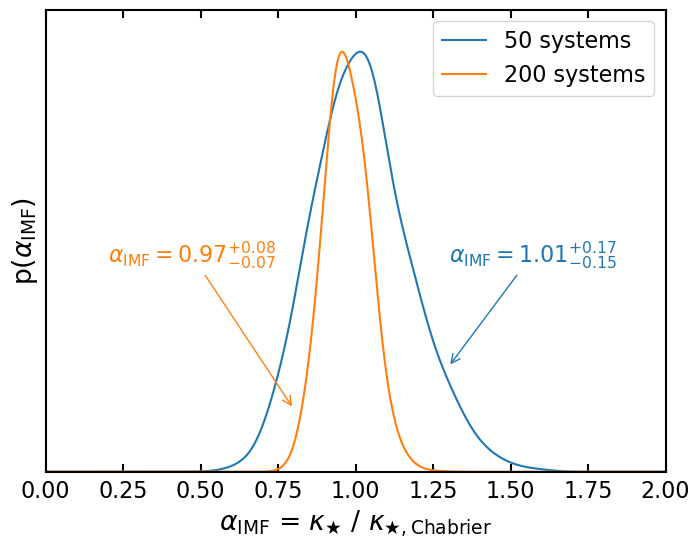}
    \includegraphics[width=\columnwidth]{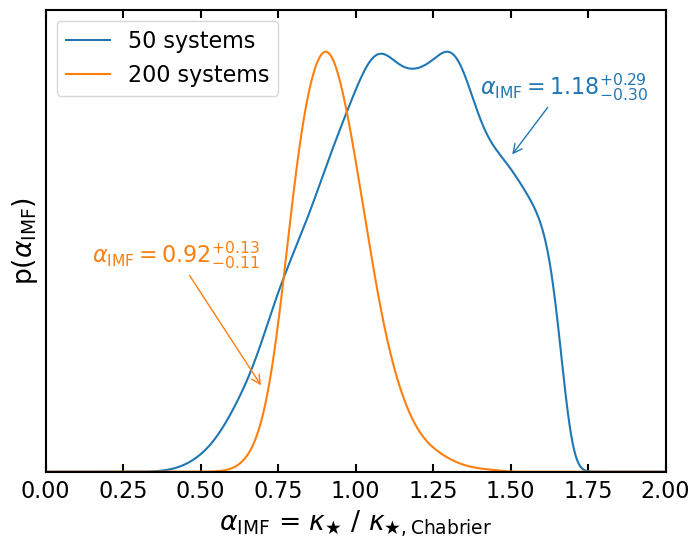}
    \caption{Posteriors of $\aimf$ for 50 or 200 glSNe Ia systems, assuming an uncertainty for the intrinsic source brightness of 0.1 mag (top) or 0.3 mag (bottom). The true value is $\aimf=1$ . In the case of larger uncertainty for the intrinsic source brightness, the sharp cutoff near $\aimf\approx 1.75$ is because $\kappa_\star$ must be $\leq\kappa$.}
    \label{fig:aimf_posteriors}
\end{figure}

We can therefore directly evaluate the simpler marginalized posterior \begin{equation}
    p(\aimf | D) = \int \left(\prod_jp(D| \aimf, m_\text{s})\right)p(m_\text{s})\odif{m_\text{s}}
\end{equation}
for each system. The product of $p(\aimf|D)$ for all systems then gives us our final constraint on $\aimf$. Implicit here is an assumption that the lenses in the sample all have the same IMF, metallicity, stellar history, and age, which allows us to combine multiple systems together. While true for our simulations as we merely used a constant $\Upsilon_\star$ for all systems in the catalog, this will of course not be true in general. Our final posterior for $\aimf$ from combining all our mock systems can be thought of as demonstrating the constraining power possible on the average $\aimf$ of a sample of lenses which are sufficiently similar in regards to stellar population.

We note that we are limited by microlensing simulations to sampling the parameter space in discrete values of $\Delta s$. These do not necessarily correspond to the same discrete values of $\aimf$ for each image within a system\footnote{Especially for the saddlepoint images where $\kappa$ is larger and a small step in $\Delta s$ is a larger step in $\aimf$ than for the images which are minima.}, which almost certainly introduces some amount of systematics into the results. We interpolate the microlensing likelihoods using kernel density estimators before multiplying them to get the final likelihood. 

\begin{figure*}
    \centering
    \includegraphics[width=2\columnwidth]{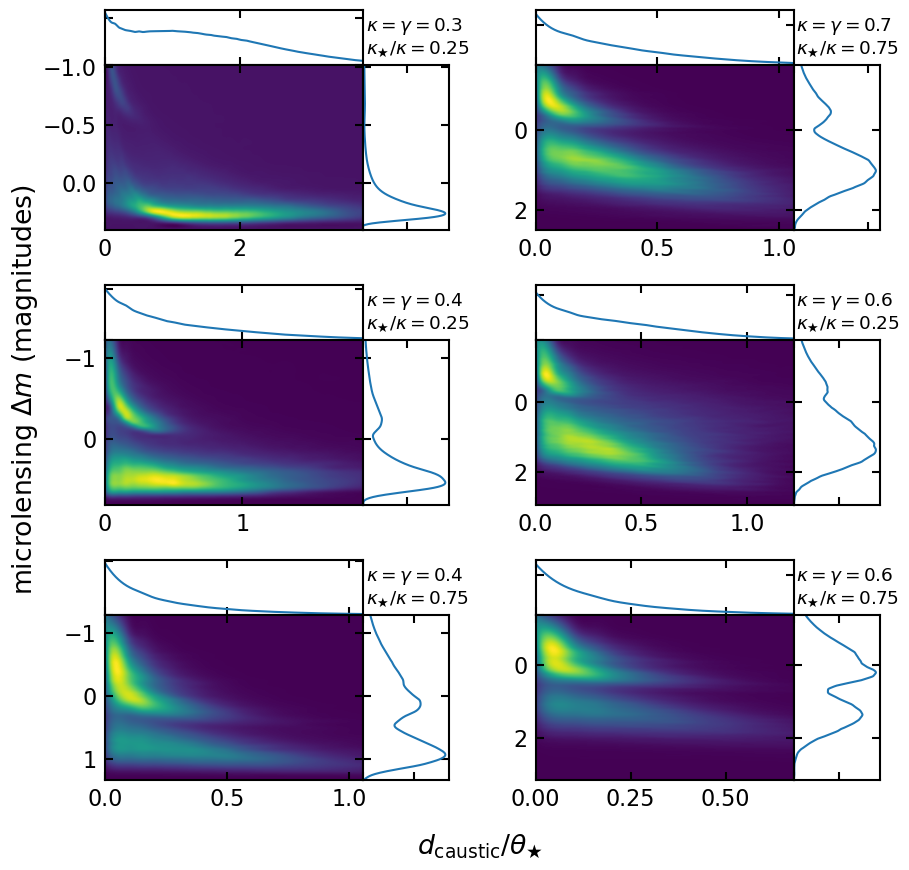}
    \caption{Joint probability distribution  of the magnification and the distance to the nearest caustic for various $\kappa=\gamma$ and $\kappa_\star$ values. Distances are in units of the Einstein radius of the microlenses $\theta_\star\propto\sqrt{m_\star}$. Note the different axes scales for each subplot, particularly for $d_{\text{caustic}}$.}
    \label{fig:d_caustic_mags}
\end{figure*}

\subsection{Results}

We initially test how well we recover $\aimf$ with 50 simulated glSNe Ia systems, which is roughly the number estimated to be found in $\sim 2-3$ years of LSST observations \citep{2023MNRAS.526.4296S, 2024MNRAS.531.3509A}. This is also a realistic number to be observationally followed up in $\sim$5 years for $H_0$ studies; our measurements are unlikely to require any extra resources on top of what is needed for time delay cosmography. We also examine how results improve with a larger number (200) of glSNe Ia. Figure \ref{fig:aimf_posteriors} shows our results when the error on the intrinsic source brightness is taken to be 0.1 or 0.3 mag. In the case of 50 systems and a 0.1 mag error on the intrinsic source brightness, $\aimf$ is recovered to within $\sim15\%$; in all cases, the true value of $\aimf=1$ is recovered within the one sigma bounds. We note once more that the prior on $\aimf$ is uniform between 0 and a physical upper bound that $\kappa_\star$ can never exceed $\kappa$. 

Since lenses are typically passive elliptical galaxies, a Salpeter IMF would have a value of $\aimf\approx1.6-1.7$ (depending on age and metallically, \citealp{2012Natur.484..485C, 2018MNRAS.475..757B}), which means that 50 glSNe Ia should be able to discriminate between a Salpeter and Chabrier IMF at the $\sim 4\sigma$ level if the error on the intrinsic source brightness is 0.1 mag. Increasing the uncertainty on the source brightness to 0.3 mag requires $\sim$200 systems to make similar constraints. Since $\aimf$ is equivalent to $\Upsilon_\star$, this means that a population of glSNe Ia should be able to constrain $\Upsilon_\star$ to $\sim 15\%$. Some additional, but likely subdominant, errors will also come from propagating measurements of the lens light and total lens galaxy mass.

\section{Constraining the mean microlens mass}
\label{sec:constraining_aimf_m}

In addition to constraining the stellar mass fraction at the image locations, light curve level data are sensitive to the mean mass of the microlenses. Some microlensed light curves will experience caustic crossing events which cause drastic changes in magnification. Due to the scaling of the Einstein radius of the microlenses with their mass, the presence or absence of caustic crossings informs us about the length scale involved so long as we know about the relative size of the supernovae. One can get the expansion velocity from spectral lines, providing a prior on the source size \citep{1974ApJ...193...27K, 1996ApJ...466..911E, 2005A&A...439..671D, 2012MNRAS.419.2783T}. For simplicity, we assume that we know the supernova size perfectly and that the time of caustic crossing is known to within a day from the observed light curves. We are then interested in adding one additional parameter back to the problem, the microlens mass $m_\star$, to constrain along with $\aimf$. 

\subsection{Likelihood simulations}

\begin{figure*}
    \centering
    \includegraphics[width=2\columnwidth]{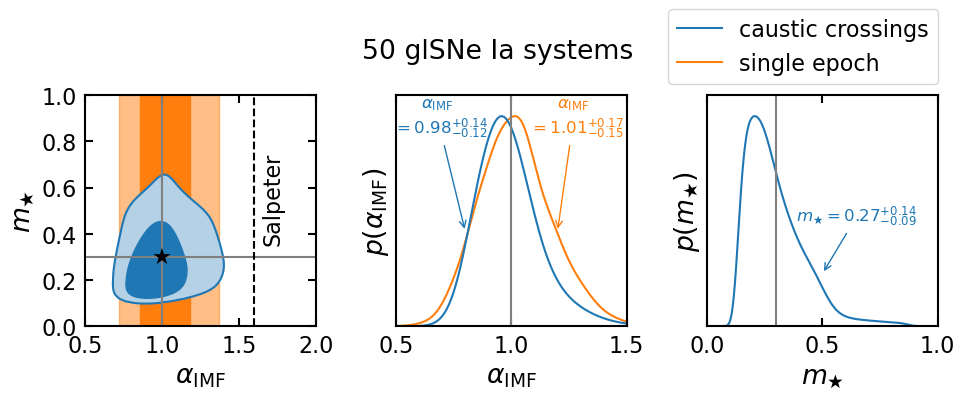}
    \caption{Joint posterior of $\aimf$ and $m_\star$ for 50 glSNe Ia systems, assuming a 0.1 mag error on the intrinsic source brightness, a uniform prior on $\aimf$ such that $\kappa_\star\leq\kappa$, and a uniform prior between 0 and 1 $M_\odot$ for $m_\star$. Solid and opaque contours denote the 68\% and 95\% confidence intervals respectively. The black dashed line marks the approximate value of $\aimf$ for a Salpeter IMF. We include the constraints from Figure~\ref{fig:aimf_posteriors} (single epoch) for comparison. See Figure~\ref{fig:aimf_mass_posteriors_vs_error} for constraints with more systems and different errors on the source brightness.}
    \label{fig:aimf_mass_posteriors}
\end{figure*}

We use a GPU implementation of \citeauthor{1990A&A...236..311W}'s \citeyearpar{1990A&A...236..311W} method to calculate the critical curves of the microlenses \inprepp{Weisenbach}, from which we can locate the caustics and create a map of the number of microminima \citep{1992A&A...258..591W, 2003ApJ...583..575G}, or equivalently the number of caustic crossings (NCC). While convolving the magnification map with the uniform disk profile of the source gives a light curve, convolving the NCC map with circular min/max filters of the same size gives two curves which together provide the location of the first caustic crossing (or whether no caustic crossing is present). 

From the NCC map we can also determine, for any pixel, the distance to the nearest caustic using a Euclidean distance transform \citep{2008ACMCS...40....F}, shown in Figure~\ref{fig:d_caustic_delta_m} (top left). Combining the magnification and NCC maps, we have a two dimensional distribution of magnification and distance to nearest caustic (Figure~\ref{fig:d_caustic_mags}), which replaces the microlensing magnification distribution in the likelihood for the joint probability of $\aimf$ and $m_\star$. 

\subsection{Joint constraints}

We jointly constrain $\aimf$ and $m_\star$ using both the observed magnification and the presence (or absence) of caustic crossings. This is similar to quasar studies which have jointly constrained the size of the quasar accretion disk and the microlens mass \citep{2017ApJ...836L..18M, 2019ApJ...885...75J}. The Bayesian framework is that discussed in Section~\ref{sec:constraining_aimf}, but we add back in the microlens mass $m_\star$ as a parameter and the time of caustic crossing as additional data, and use the joint probability distribution of the microlensing magnification and the distance to the nearest caustic in our likelihood. 

We show in Figure~\ref{fig:aimf_mass_posteriors} the constraints for 50 systems, assuming a 0.1 mag uncertainty on the intrinsic source brightness and a uniform prior on $m_\star$ between 0 and 1 $M_\odot$. As before, $\aimf$ is constrained to within $\sim 15\%$. The constraints on $m_\star$ are more broad, but the input of $0.3M_\odot$ is well recovered. 

If the number of supernovae systems is increased from 50 to 200 while maintaining a 0.1 mag uncertainty on the source brightness, the constraints on $\aimf$ improve from $0.98^{+0.14}_{-0.12}$ to $1.01^{+0.07}_{-0.06}$, while the constraints on $m_\star$ improve from $0.27^{+0.14}_{-0.09}$ to $0.27^{+0.07}_{-0.06}$. Increasing the source brightness uncertainty to 0.3 mag while maintaining 200 systems changes the constraints to $\aimf=1.12^{+0.11}_{-0.10}$ and $m_\star=0.27^{+0.10}_{-0.06}$ (see also Figure~\ref{fig:aimf_mass_posteriors_vs_error}). In general, the constraints on $\aimf$ appear more sensitive to changes in source brightness uncertainties than the constraints on $m_\star$.

\section{Caveats}
\label{sec:caveats}

We assume that the macro-model for the lens is known precisely. This will, in general, not be true. In particular, the mass-sheet degeneracy \citep{1985ApJ...289L...1F, 2014A&A...564A.103S} introduces uncertainties into the convergence which are important for microlensing, as $\kappa$ is a necessary constraint for microlensing measurements. However, the uncertainties on $\kappa$ that arise from the mass-sheet degeneracy are unlikely to lead to changes in the magnification at the $0.2-0.3$ mag level as considered here; a few percent uncertainty on the mass-sheet degeneracy parameter implies a subdominant ($<10\%$) uncertainty on the magnification. Lens modeling uncertainties for $\kappa$ from fitting overly simple lens models and neglecting substructures are likely to be larger sources of error, though our consideration of 0.3 mag errors on the magnifications is likely still a pessimistic scenario. Furthermore, either the Hubble constant and measurements of the time delays between images or lens velocity dispersions can be used to constrain the macro-model. In practice therefore, a true analysis would require marginalizing over values of $\kappa$ and $\gamma$ in the lens macro-model, which may significantly affect the microlensing likelihoods in some regions of parameter space \citep{2014MNRAS.445.1223V}. This would require image level data and lens modeling which, while outside the scope of this work, will be important to implement for future glSNe discoveries. 

Millilensing poses a particular challenge to our ability to measure $\kappa_\star$, as some portion of the flux ratio anomalies could be due to millilensing by dark matter subhalos as opposed to microlensing by stars. While the narrow-line emission regions of quasars are large enough to be insensitive to microlensing yet small enough to be sensitive to millilensing \citep{2003MNRAS.339..607M, 2014MNRAS.442.2434N}, allowing one to disentangle the contributions of micro- and millilensing and constrain the subhalo mass function \citep{2020MNRAS.491.6077G}, glSNe do not offer such an opportunity. Joint modeling of glSNe Ia flux with subhalo magnifications \citep[e.g.][]{2025ApJ...980..172L} in addition to microlensing magnifications may constrain both, albeit at the requirement of more systems.

As previously mentioned, the majority of systems in the catalog of mocks used here are doubly imaged supernovae. The lens models for such systems are typically under-constrained if only the images of the supernovae are seen. Having a lensed host galaxy makes the modeling easier, but only $\sim50\%$ of glSNe will have a lensed host with a smaller ($\sim25\%$) fraction having a lensed host which is bright enough to observe \citep{2024MNRAS.535.2523S}. Quadruply lensed systems have better macromodel constraints on $\kappa$ and $\gamma$, and close pairs of fold images are ideal for constraining $\Upsilon_\star$ \citep{2002ApJ...580..685S}, but they consist of a smaller population. The case of a 0.3 mag uncertainty on the intrinsic source brightness can be considered as an imperfect proxy for inflating the errors from uncertain lens models, highlighting the larger number of systems that might be required for precise constraints.

While we ignore the effect of the size of the source on magnifications, a better analysis would consider the joint likelihood of microlensing magnification, $d_{\text{caustic}}$, and source size. However, the magnification of a supernova that does not cross a microcaustic is essentially constant over its lifetime, changing nothing in our results. A supernova that does cross a microcaustic will experience a change in magnification which may help constrain $\aimf$ and $m_\star$ better, but the limited number of such supernovae and the fact that our posteriors still capture the true values suggests that $d_{\text{caustic}}$ provides enough information and further improvements would be minimal. 

Our analyses lack in one area which will be important to explore in the future. While a given IMF determines the stellar mass fraction at the image positions, it also determines the PDMF. The dependence of microlensing on the mass spectrum is usually ignored, largely due to the fact that it is difficult to manifest in magnification distributions. It was originally conjectured to bear little to no role \citep{1992ApJ...386...19W, 1995MNRAS.276..103L, 1996MNRAS.283..225L}, though that was later disproven by \citet{2004ApJ...613...77S}. The dependence is minimal in microlensed quasars for physical reasons \citep{2006ApJ...645..835L} which are also relevant for supernovae: at a fixed size the source is blind to masses below some cutoff. Microlensing studies must therefore be careful to state the mass range for which results about the typical mass are valid when incorporating constraints on the source size.


Convolutions of the microlensing magnification maps with a source profile wash out the contributions of low mass microlenses, making them act as a smooth component indistinguishable from dark matter. Although we use a single mass for all of the microlenses for our simulations and subsequent constraints, a proper analysis of real data would need to consider a spectrum of masses given by the PDMF. This widens the parameter space, as one must then worry about the upper and lower mass cutoffs as well as the slope (and other possible parameters governing the shape) of the mass spectrum. While such simulations are feasible with current computational resources for microlensing, they do multiply the necessary time -- which already took $\sim$weeks to generate the necessary microlensing likelihoods for our efforts. The development of (semi)analytic expressions or approximations for the microlensing likelihoods are desirable and would make future analyses more tractable.

\section{Discussion and Conclusions}
\label{sec:conclusions}

The present day stellar mass in galaxies is dependent on both the initial stellar population and subsequent star formation history. Estimates of stellar mass from SPS can be uncertain by factors of $\sim 2$ or more, depending on stellar evolution uncertainties, initial galactic metallicity distributions, and assumptions regarding the IMF \citep{2009ApJ...699..486C}. Robust independent measurements of stellar mass provide a means for testing SPS models and probing IMF dependencies on galactic parameters such as metallicity, velocity dispersion, and redshift. Contrary to methods which infer stellar mass from measured light, gravitational microlensing provides a unique way to directly probe the mass of compact objects in lensing galaxies using their gravitational effect. Microlensing can simultaneously measure the fraction of mass in compact objects and the mean mass of those objects, directly providing constraints on the lens galaxy PDMF and hence indirectly on the IMF. 

In this work, we have used a simulated sample of glSNe Ia to examine how well we can jointly constrain an IMF mismatch parameter $\aimf$ (equivalent to a stellar mass-to-light ratio $\Upsilon_\star$) and the mean stellar mass $m_\star$ of the microlenses, by exploiting observed magnifications of glSNe Ia and the timings of caustic crossings as the SN expands. With a sample of 50 glSNe Ia systems, we consistently recover our input $\aimf$ and $m_\star$ values within the 68\% confidence intervals of the posteriors. The value of $\aimf$ is recovered to within $\sim 15\%$ if the uncertainty on the intrinsic brightness of the source is 0.1 mag, while $m_\star$ is constrained to within $\sim 50\%$. While we used total stellar mass estimates in our simulations from strong lensing fits \citep{2010ApJ...724..511A} with a Chabrier IMF, a Salpeter IMF would differ by a factor of $\sim1.6$ \citep{2017ApJ...841...68V, 2024ApJ...973L..32V}, making our microlensing constraints a $\sim4\sigma$ discriminator between the two. 

The mean stellar mass is not constrained enough to differentiate between IMFs, even when the number of systems is increased to 200. If the first image is discovered early enough, gravitational lensing allows us to predict when and where the second image will appear \citep{2020MNRAS.495.4622F}. Early observations of glSNe (when the SN photosphere is physically small) should be sensitive to the low mass end of the PDMF. While our work has assumed a single mass microlens population, our approach should generalize to a mass spectrum and there is hope for better constraints if some of our simplifying assumptions are removed; \cite{2001MNRAS.320...21W} showed that light curve variability statistics for lensed quasars can be inverted to obtain the microlens mass spectrum, a promising result which should also be explored for glSNe. 

While the exact precision of our results should be viewed cautiously given the simplifications we made regarding lens modeling, our results suggest that even with minimal priors glSNe Ia have excellent sensitivity with which to probe the stellar mass content of lens galaxies and constrain their average IMF. Even if we have underestimated errors by a factor of 2, microlensing of 50 glSNe Ia would still be a $2\sigma$ discriminator between a Chabrier and Salpeter-like IMF in lens galaxies. 

Multispectral observations of glSNe should provide slightly improved constraints from what we expect, as the measurement of $\kappa_\star$ does not greatly depend on the filter used since the supernova size does not change by a large factor in different wavelengths. Once the supernova fades away, the value of $\Upsilon_\star$ in various filters can then be determined -- a quantity which is slightly harder to measure for lensed quasars, as they are always present and require deblending and simultaneous modeling with the lens galaxy light. Conversely, simultaneous modeling of the supernova half-light radii in various filters with the effects of microlensing may help provide additional constraints on theoretical supernova explosion models \citep{2018ApJ...855...22G, 2019A&A...631A.161H, 2021A&A...646A.110H, 2021A&A...653A..29B}. 

We assumed a constant mass-to-light ratio in our simulations, but there is increasing evidence for radial gradients in $\Upsilon_\star$ from spectroscopy \citep{2017ApJ...841...68V} and lensing and dynamics \citep{2018MNRAS.476..133O, 2018Sci...360.1342C, 2018MNRAS.481..164S}. It should be possible to extend our approach to probe mass-to-light gradients in lens galaxies given the different radii at which lensed macroimages form, though such measurements may be more difficult due to the demagnification of the saddlepoint counterimages -- which also have an increased likelihood of yet further microlensing demagnification \citep{2002ApJ...580..685S}. 



In the next decade glSNe will open a new window onto the population of stars and compact objects in galaxies. With modest investment in follow-up of the LSST glSNe Ia population we should be able to discriminate between a Chabrier and Salpeter IMF with $\sim 4 \sigma$ confidence, completely independent of other methods and without the need to assume the dark matter profile of the lens galaxy.

\section*{Acknowledgements}
We would like to thank Scott Gaudi and Enrique Gaztanaga for raising the question of what one can learn about the lens galaxy and microlensing from glSNe Ia if cosmology is known. We thank Paul Schechter for his careful reading and comments. We would also like to thank the referee, Alessandro Sonnenfeld, for his careful reading and feedback, which greatly improved the presentation and quality of this work.

Numerical computations were done on the Sciama High Performance Compute (HPC) cluster which is supported by the ICG, SEPNet, and the University of Portsmouth. 

This work has received funding from the European Research Council (ERC) under the European Union’s Horizon 2020 research and innovation programme (LensEra: grant agreement No. 945536). TC is funded by the Royal Society through a University Research Fellowship. For the purpose of open access, the authors have applied a Creative Commons Attribution (CC BY) license to any Author Accepted Manuscript version arising.

\section*{Data Availability}

The simulated data from this work will be made available upon reasonable request to the corresponding author. 



\bibliographystyle{mnras}
\bibliography{bibliography.bib} 




\appendix

\section{Joint constraints on \texorpdfstring{$\aimf$}{α\textunderscore IMF} and \texorpdfstring{\lowercase{$m_\star$}}{m\textunderscore ⭑}}

\begin{figure*}
    \centering
    \includegraphics[width=2\columnwidth]{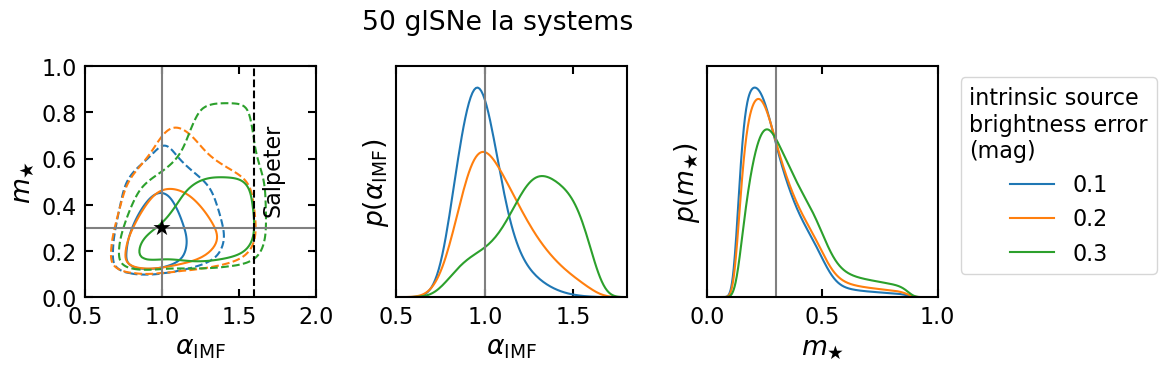}\\
    \vspace{20pt}
    \includegraphics[width=2\columnwidth]{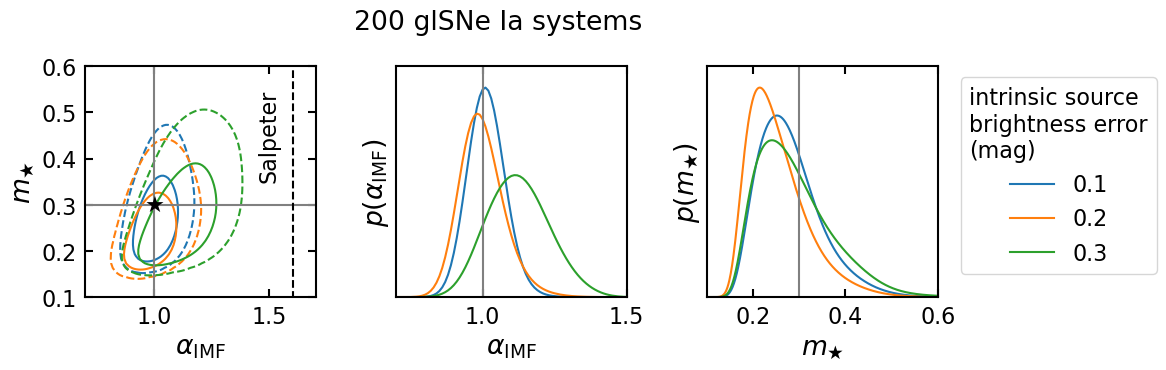}
    \caption{Joint posterior of $\aimf$ and $m_\star$ for 50 or 200 glSNe Ia systems, assuming various errors on the intrinsic source brightness, a uniform prior on $\aimf$ such that $\kappa_\star\leq\kappa$, and a uniform prior between 0 and 1 $M_\odot$ for $m_\star$. Solid and dashed contours denote the 68\% and 95\% confidence intervals respectively, while the black dashed line marks the approximate value of $\aimf$ for a Salpeter IMF. Note the different axes ranges between the top and bottom plots.}
    \label{fig:aimf_mass_posteriors_vs_error}
\end{figure*}

In Figure~\ref{fig:aimf_mass_posteriors_vs_error}, we show joint constraints on $\aimf$ and $m_\star$ for either 50 or 200 glSNe Ia systems, with errors on the intrinsic source brightness varying from 0.1 to 0.3 mag. The true values $\aimf=1$ and $m_\star=0.3M_\odot$ lie within or near the 68\% confidence intervals for the joint posteriors, though $\aimf$ lies slightly outside the 68\% interval of its marginalized posterior when the error on the source brightness is large. This is likely due to the multimodal nature of the microlensing likelihoods playing a larger role when the prior on the source brightness is wider.

\bsp	
\label{lastpage}
\end{document}